\documentclass{appolb}
\usepackage{epsfig}

\def\bge{\begin{equation}}
\def\ene{\end{equation}}
\def\bg{\begin{eqnarray}}
\def\en{\end{eqnarray}}

\def\bge{\begin{equation}}
\def\ene{\end{equation}}
\def\bg{\begin{eqnarray}}
\def\en{\end{eqnarray}}

\begin{document}
\title{
THE COSMOLOGICAL CONSTANT \\
AND STABILITY OF THE HIGGS VACUUM
\thanks{Presented at the Jagiellonian Symposium on Fundamental and Applied Subatomic Physics, Cracow, June 7-12 2015.}
}
\author{Steven D. Bass 
\address{Stefan Meyer Institute for Subatomic Physics, \\
Austrian Academy of Sciences,
Boltzmanngasse 3, 1090 Vienna, Austria}
}
\maketitle
\begin{abstract}
\noindent
We discuss the cosmological constant puzzle and possible
connections to the (meta-)stability of the Higgs vacuum
suggested by recent LHC results. 
A possible explanation involves new critical 
phenomena in the ultraviolet, close to the Planck scale.
\end{abstract}

\section{Introduction}

The cosmological constant puzzle connects particle physics 
and cosmology.
The accelerating expansion of the Universe is interpreted
within Einstein's theory of General Relativity as driven 
by a small positive cosmological constant 
or vacuum energy density perceived by gravitational interactions called dark energy,
for reviews see Refs.~\cite{weinberg,bassmpla}.
Understanding this vacuum energy is an important challenge 
for theory and connects the Universe on cosmological scales 
(the very large) with quantum field theory and subatomic physics (the very small).

Current observations point to an energy budget of 
the Universe where just 5\% is composed of matter built
from quarks and leptons, 
26\% involves dark matter (possibly made of new elementary particles) and 69\% is dark energy~\cite{planck15a}.
The vacuum energy density extracted from astrophysics is 
\begin{equation}
\rho_{\rm vac} = \mu_{\rm vac}^4 \sim (0.002 \ {\rm eV})^4 .
\end{equation}
The scale $\mu_{\rm vac}$ is similar to the value
we expect for the light neutrino mass whereas $\rho_{\rm vac}$
is 10$^{56}$ times smaller than the value expected from the classical Higgs potential of Standard Model particle physics which also comes with the opposite negative sign.
What might dark energy be telling us about 
the intersection of particle physics and gravitation?

General Relativity and the particle physics Standard Model 
work excellently everywhere they have been tested in experiments. 
The Higgs boson discovered at 
the LHC \cite{higgs:lhc} is consistent 
with Standard Model expectations \cite{guido}.
It is an open question whether at a deeper level 
this boson is elementary or of dynamical origin. 
Results from the LHC experiments ATLAS, CMS and LHCb are 
in good agreement with the Standard Model with (so far) 
no evidence of new physics.
Precision measurements of the electron 
electric dipole moment are consistent with zero and constrain possible new sources of CP violation from beyond the Standard Model up to scales similar to or larger than those probed at the LHC \cite{acme}.
Whilst the Standard Model has proved very successful 
everywhere it has been tested we know that some extra 
physics is needed to explain the very small neutrino masses, 
the baryon asymmetry and strong CP problems as well as 
dark matter and inflation. The scale of this new physics 
is as yet unknown and not yet given by experiments.

\section{The cosmological constant and particle physics}

Interestingly,
the dark energy scale $\mu_{\rm vac} \sim 0.002 \ {\rm eV}$ 
is similar to the value that we expect for the light neutrino 
mass~[7--9]
taking
the normal hierarchy of neutrino masses~\cite{neutrinomass}.
One observes 
the phenomenological relation~\cite{bassmpla,bass:acta}
\begin{equation}
\mu_{\rm vac} \sim m_{\nu} \sim \Lambda_{\rm ew}^2/M \ ,
\end{equation}
where $\Lambda_{\rm ew}$ is the electroweak scale and
$M \sim 3 \times 10^{16}$ GeV is logarithmically close 
to the Planck mass 
$M_{\rm Pl} \sim 1.2 \times 10^{19}$ GeV 
and typical of the scale that 
appears in Grand Unified Theories and in the see-saw 
mechanism for neutrino mass generation~\cite{seesaw}.
If taken literally Eq.(2) connects neutrino physics, 
Higgs phenomena in electroweak symmetry breaking and 
dark energy to a new high mass scale which needs to be understood.
The gauge bosons in the Standard Model which have 
a mass through the Higgs mechanism are also the gauge bosons 
which couple to the neutrino.

The formula Eq.(2) was also suggested 
by Bjorken~\cite{bjcosmo} 
without connection to neutrinos
in the ``gaugeless limit'' of the Standard Model
with composite or emergent gauge bosons being born at a
large mass scale $M$ and no or only very small
coupling to new physics between the electroweak and ultraviolet mass scales.

There are theoretical hints that the large mass scale 
in Eq.(2) might perhaps be associated with dynamical 
symmetry breaking, see below.
The see-saw formula Eq.(2), 
if evidence of deeper physics, suggests that the 
cosmological constant puzzle and the electroweak 
hierarchy problem 
might be connected and perhaps
be resolved at a scale close to the 
Planck scale rather than around the TeV scale.

Here one finds a possible connection with LHC results.
With the values of the top-quark and Higgs boson masses
measured at the LHC, the Standard Model works as a 
consistent perturbative theory up to very high scales.
Renormalisation group (RG) calculations reveal 
that the Standard Model Higgs vacuum 
with no coupling to new interactions 
sits close to the border of being stable and metastable 
(with half-life much greater than the present age of the Universe)
[13--20].
An unstable vacuum would require coupling to some 
new interaction above the electroweak scale. 
Vacuum stability is very sensitive to the exact values of 
the Higgs and top-quark masses and technical details in calculating 
$\overline{\rm MS}$ parameters in terms of physical ones
and how one should include tadpole diagrams to be consistent 
with gauge invariance. 
The key issue is that the $\beta$ function for the Higgs 
four-boson self-coupling $\lambda$ has a zero 
around $10^{17}$ GeV, close to the Planck mass,
and when 
(if at all) this coupling $\lambda$ 
crosses zero,
perhaps around 
$10^{10}$--$10^{12}$ GeV \ \cite{degrassi}
or, for a stable vacuum, not at all \cite{masina,fredj1}. 
With modest changes in the top-quark and Higgs masses
(increased top mass and/or reduced Higgs mass) 
the Standard Model vacuum would become unstable.
In a recent calculation 
Bednyakov et al. \cite{kniehl} find 
the value of the top quark mass for the vacuum 
to be stable all the way up to the Planck mass to be
within $1.3 \sigma$ of the Monte Carlo mass quoted by the 
Particle Data Group.
With the vacuum either stable up to the Planck scale or 
at the border of stable and metastable, then some critical 
process might be at work in the 
ultraviolet~\cite{degrassi,fredj1}.

Ideas connecting the cosmological constant and Higgs-mass hierarchy problem to new critical phenomena near the Planck 
scale are discussed in 
Refs.~\cite{bassmpla,bass:acta,bjcosmo,fredj1,fredj2}.
Perhaps the Standard Model is itself (in part) 
emergent as the long range tail of a critical 
system that exists close to the Planck scale 
and there is no new scale between the electroweak 
scale and some very high scale close to the Planck mass?

In parallel to RG discussions of vacuum (meta-)stability, 
the RG behaviour of the perturbative coefficient of 
the quadratically divergent counterterm 
for the Higgs mass squared is also interesting.
This coefficient crossing zero \cite{veltman} in 
the ultraviolet would trigger a first order phase transition restoring electroweak symmetry \cite{fredj1}.
Whether this crossing transition happens above or 
below the Planck scale depends strongly on the 
value of the top-quark mass and 
matching between the $\overline{\rm MS}$ and 
physical parameters~\cite{degrassi,hamada,fredj1,masinaq}.
In the calculation of Jegerlehner~\cite{fredj1} 
with a stable vacuum the crossing transition takes place around $10^{16}$ GeV, close to the mass scale $M$ in Eq.(2).

We next consider how Eq.(2) might be understood, treating 
the chirality of the neutrino by analogy as an Ising-like ``spin'' degree of freedom that becomes active near the 
Planck scale \cite{bass:acta}.
Analogies between quantum field theories and condensed matter and statistical systems have often played an important role in 
motivating ideas in particle physics.
The ground state of the Ising model exhibits spontaneous magnetisation where all the spins line up and the internal 
energy per spin and the free energy density of the spin 
system go to zero with corrections dampened by a strong suppression exponential factor, 
{\it viz.} $e^{- \beta J}$ where $\beta$ is Boltzmann's 
constant and $J$ is the scale of the Ising interaction.
In mappings between statistical mechanics and quantum field theory the free energy density for the statistical 
``spin'' system plays the role of the vacuum energy 
density in quantum field theory \cite{kogut} suggesting possible application to the cosmological constant puzzle.
For an Ising system with no external magnetic field
the free energy density is equal to minus the pressure.
The model equation of state looks like a vacuum energy 
term in Einstein's equations of General Relativity, 
proportional to the metric tensor $g_{\mu \nu}$.
Taking the scale of the Ising interaction $J \sim M$
close to the Planck mass and 
coupling the ``neutrino'' spins to gauge fields
it is plausible that the gauge bosons which couple 
to the ``neutrino'' non-perturbatively acquire mass
in the ground state~\cite{bassmpla,bass:acta}.
That is, they are in a Higgs phase.
Further, the lowest energy eigenvalue characterising 
the free energy of the combined ``spin''-gauge system 
then looks like the see-saw formula, Eq.(2),
with large mass scale $M \sim 10^{16}$ GeV.

If the gauge symmetries of the Standard Model are emergent,
this differs from the paradigm of unification with maximum 
symmetry at the highest possible energies with a unification big gauge group spontaneously broken through various Higgs condensates to the Standard Model, with each new condensate introducing an extra large contribution to the vacuum energy and the cosmological constant.
With emergent gauge fields [18, 25--28] Lorentz invariance 
is expected to be (spontaneously) broken or emergent close 
to the critical mass scale in the ultraviolet \cite{bj2001}.
Bjorken has argued that any violations of Lorentz and 
gauge symmetries in the emergence scenario might appear 
with coefficient suppressed by powers of the cosmological 
constant scale divided by the large scale 
$M$ close to the Planck mass \cite{bj2001,bj2010}, 
thus vanishing in the limit of vanishing dark energy.
For emergent QED a preferred reference frame is naturally identified as the frame for which the cosmic microwave background is locally at rest.
Non-renormalisable contributions from high dimensional
operators would be proportional to powers of energy
divided by $M$ and are very much suppressed 
much below the Planck scale \cite{fredj1}.
In the emergence scenario fundamental symmetries like gauge 
and Lorentz invariance would ``dissolve" in the ultraviolet 
and could be manifest as infrared attractive fixed points 
of the renormalisation group behaviour of a larger class of 
possible theories \cite{nielsen}.
Perhaps the gauge theories of particle physics and also 
General Relativity are effective theories with characteristic energy of order the Planck scale~\cite{weinberg09}.

\section{Conclusions}

The cosmological constant puzzle continues to challenge 
our understanding of fundamental physics.
Why is the dark energy density finite, positive and 
so very small?  
Might the value of the cosmological constant and 
electroweak symmetry breaking be related, 
perhaps with common origin connected to new critical 
phenomena in the ultraviolet close to the Planck scale?
The cosmological constant puzzle promises to teach us a 
great deal about the intersection of particle physics and
gravitation.

\section*{\bf Acknowledgements}

I thank Pawel Moskal for the invitation to this stimulating meeting in the beautiful surroundings of the Collegium Maius.

\end{document}